\documentclass[preprint,aps,draft]{revtex4}

\usepackage{bm}

\begin{document}

\preprint{IZTECH-P05-2004}

\title{A symmetry for vanishing cosmological constant in an extra 
dimensional toy model}

\author{Recai Erdem}
\email{recaierdem@iyte.edu.tr}
\affiliation{Department of Physics,
{\.{I}}zmir Institute of Technology \\ 
G{\"{u}}lbah{\c{c}}e K{\"{o}}y{\"{u}}, Urla, {\.{I}}zmir 35430, 
Turkey} 

\date{\today}

\begin{abstract}
We introduce a symmetry principle that forbids a bulk cosmological 
constant in 
six and ten dimensions. Then the symmetry is extended in six dimensions so 
that it insures absence of 4-dimensional cosmological constant induced by 
the six dimensional curvature scalar, at least, for a class of metrics. 
A small cosmological constant may be induced in this scheme by breaking 
of the symmetry by a small amount.
\end{abstract}

\maketitle

Cosmological constant problem is a long standing problem \cite{Hist}. The 
problem can be stated as
the huge discrepancy between the observational and the theoretically 
expected values of the cosmological constant \cite{Wein} and the lack of 
understanding 
of its extremely small value \cite{PDG}.  Numerous schemes, to solve this 
problem, 
range from 
the models which employ supersymmetry, supergravity, superstrings, 
anthropic principles, modified general relativity, self-tuning 
mechanisms, quantum cosmology, extra dimensions, and combinations of these 
ideas \cite{Wein,dynamical, small,infinite,self}. 
Although they shed some light on the direction of the solution of this 
problem, they have not given a wholly satisfactory, widely accepted answer 
to this question.  Among these attempts extra dimensional models become 
more popular because they give model builders more 
flexibility \cite{small,infinite,self}. This is mainly due to the 
fact the no-go theorem of Weinberg \cite{Wein} is intrinsically 
four dimensional; for example, the equations of motion for a field 
constant in 4-dimensions may contain a contribution from extra 
dimensional kinetic term in the Lagrangian hence making the Weinberg's 
argument non-applicable in higher dimensions. Moreover the models where a 
four dimensional space is embedded in a 
higher dimensional space may have striking differences. For example four 
dimensional world may be embedded in extra dimensions in such a way that 
the 4-dimensional 
brane remains flat under energy density changes on the brane through the 
counter balance of the curvature due to the extra dimensions and 
the brane tension \cite{self}. However these models, although appealing, 
at present have some technical problems such as need for large extra 
dimensions which may be in conflict with astronomical data, fine tuning, 
technicalities with quantum loop corrections, severe restrictions on 
dilaton-brane couplings \cite{burgess}.  So additional insight on 
the cosmological constant in extra dimensions may be useful. 
In this paper we study the implications of a symmetry, similar to scale 
invariance with a complex scale factor, on the  cosmological constant. 
In fact it seems that such a symmetry principle was also noticed by 't Hooft 
(though unpublished) \cite{nobbenhuis}. We find that this symmetry 
forbids a non-zero bulk cosmological constant in 6 and 10 dimensions.  
We consider the 6 dimensional case in this paper. We extend the symmetry 
so that the contribution to the cosmological constant due to the extra 
dimensional curvature scalar vanishes as well. We find that breaking 
of the symmetry by a small amount may result in a small 
cosmological costant in this scheme. We also briefly discuss the 
restriction put on the form of matter Lagrangian by this symmetry.

Consider the transformation which multiplies the coordinates by the 
imaginary number $i$
\begin{equation}
    x_A \rightarrow i\, x_A~~,~~~~~~~~ A=0,1,......, D-1
\label{a1a} 
 \end{equation}
where $D$ stands for the dimension of the space. This transformation 
may be viewed as an analytic continuation followed by a rotation by 
$\frac{\pi}{2}$ in each complex plane. We impose the symmetry
\begin{equation}
g_{AB}\rightarrow g_{AB}~~~~\mbox{as}~~~~
    x_A \rightarrow i\, x_A~~,~~~~~A=0,1,......, D-1
   \label{a1b} 
 \end{equation}
Under Eq.(\ref{a1b}) the scalar curvature is multiplied by $-1$ 
\begin{equation}
R \rightarrow -R \label{a2}
\end{equation}
and
\begin{eqnarray}
~~~~~~~~~~~~~~~d^Dx~~~~&&\mbox{if}~~~D=4n \label{a3a} \\ 
d^Dx
~\rightarrow 
~~~~~~~~~~~~~~~-\,d^Dx 
~~~~&&\mbox{if}~~~D=2(2n+1) 
\label{a3b} \\
~~~~~~~~~~~~~~~
\pm\,i\,d^Dx
~~~~&&\mbox{if}~~~D=2n+1
\label{a3c} \\
&&n=0,1,2,3,............. \nonumber
\end{eqnarray}
The requirement that Einstein-Hilbert action
\begin{equation}
S_R = \frac{1}{16\pi\,G}\int \sqrt{g} \,R \,d^Dx \label{a4}
\end{equation}
should be invariant under (\ref{a1b}) selects out 
\begin{equation}
D=2(2n+1)~~~,~~~~n=0,1,2,3,....
\label{a5}
\end{equation}
In fact, in the case of exact symmetry 
the action should be invariant up to a multiplicative constant because the 
equations of motion remain the same. However, if the symmetry is broken ( 
even by a small amount which is the physical situation) then the part of 
the action respecting the symmetry must be strictly invariant since each 
constant multiplying the symmetry preserving part of the action leads to a 
different equation of motion in general after taking the symmetry breaking part into account. 
Another point worthwhile to mention is that under 
Eq.(\ref{a1b})
\begin{equation}
ds^2~\rightarrow ~-\,ds^2
\label{a6}
\end{equation}
This implies a symmetry under exchange of space-like and time-like 
intervals. The implications of this transformation need a separate study.

We notice from the equations (\ref{a2}) and 
(\ref{a3b}) that the cosmological constant term for the action
\begin{equation}
S_C = \frac{1}{16\pi\,G}\int \sqrt{g} \,\Lambda \,d^Dx \label{a7}
\end{equation}
( where $\Lambda$ is constant in $x_A$, $A=0,1,....D-1$ ) is not allowed 
by the symmetry induced by Eq.(\ref{a1a}) for $D$'s satisfying (\ref{a5}). 
Under the requirement $D\geq 4$ and $D\leq 10$ (which comes from string 
theory) the only possible dimensions allowed by the symmetry induced by 
(\ref{a1b}) are 6 and 10. 

In this paper we study the minimal case i.e. 
$D=6$ case. It is evident that an $S_R$ term is not allowed in $D=4$. On 
the other 
hand a cosmological constant term,  Eq.(\ref{a7}) is allowed in 4 
dimensions. In other words, although the invariance of the action under
(\ref{a1b}) forbids a six dimensional cosmological constant it does not 
forbid a 4-dimensional one. So a 4-dimensional cosmological constant may 
be induced through the six dimensional curvature scalar even if there is no 
contribution to it through a six dimensional bulk cosmological constant. A 
$D$-dimensional curvature scalar 
( $D>4$ ) may be written as
\begin{equation}
R = R_1(x_\mu,x_a)+R_2(x_a)~~~,~~~~~\mu=0,1,2,3~~~;~~~~a=4,5,.......,D-1 
\label{a8} 
\end{equation}
where $R_2$ is the part of the curvature scalar which is independent of 
$x_\mu$ (i.e. $R_2$ depends only on the extra dimensions) and $R_1$ is the 
part which contains $x_\mu$ -dependent and the mixed terms. A 
non-vanishing $R_2$ in general introduces a cosmological constant in 
4-dimensions. So one must impose a symmetry which eliminates $R_2$ as well 
in order to make the 4-dimensional cosmological constant zero altogether. 

The local 4-dimensional Poincare invariance in six dimensions results in 
the metric \cite{rubakov}
\begin{equation}
ds^2=\sigma(x^a)g_{\mu\nu}dx^\mu\,dx^\nu + 
g_{ab}(x^a)dx^a\,dx^b~~~,\mu\nu=0,1,2,3~~;~a,b=4,5
\label{a9}
\end{equation}
In addition to the symmetry (\ref{a1b}) we require
\begin{equation}
g_{AB}~\rightarrow~g_{AB}~~~~\mbox{as}~~~x_4\,\leftrightarrow\,x_5
\label{a10}
\end{equation}
and take
\begin{equation}
g_{44}=-g_{55} \label{a11}
\end{equation}
( which may be obtained by imposing 
$g_{ab}dx^a\,dx^b\,\rightarrow\,g_{ab}dx^a\,dx^b$ under Eq.(\ref{a1b}) ).
Under these requirements we find that the extra dimensional 
components of the Riemann tensor are zero, $R_{ab}=0$, and its 
4-dimensional part $R_{\mu\nu}$ depends only on the 4-dimensional coordinates $x^\mu$. So 
$R_2(x^a)=0$ in this case. In other words the local 4-dimensional Poincare 
invariance together with the requirements Eqs.(\ref{a10}-\ref{a11}) 
guarantee the absence of a contribution from 6-dimensional curvature scalar 
to the 4-dimensional cosmological constant.
 
We have introduced the symmetry (\ref{a1b}) to eliminate the 
6-dimensional bulk cosmological constant and the symmetry 
(\ref{a10}-\ref{a11})) to eliminate 
a possible contribution to the 4-dimensional cosmological constant from 
6-dimensional curvature scalar. Now we give some examples first to see 
the picture more clearly and then consider the case of the symmetry 
breaking by a small amount. 
First take the metric 
\begin{eqnarray}
ds^2&=&
\Omega_1^2(y)
g_{\mu\nu}(x)\,dx^{\mu}dx^\nu\,+\,\Omega_2^2(y)
\eta_{ab}\,dy^{a}dy^b 
\label{a12} \\
&&(\eta_{ab}) =\mbox{diag}(-1,1) ~~~~~~~\mu,\nu=0,1,2,3~~;~~a,b=1,2  
~~~~~~y_1=x_4~,~y_2=x_5  
\nonumber
\end{eqnarray}
where $x$ stands for the 4-dimensional coordinates and $y$ for the extra 
dimensional coordinates. Provided that the metric tensor is a smooth 
function of $y$ the curvature scalar corresponding to (\ref{a12}) 
is
\begin{eqnarray}
R\,=
\Omega_1^{-2}
g^{\mu\nu}R_{\mu\nu}
&-&\Omega_1^{-2}[
10\eta^{ab}\partial_a\partial_b(ln\Omega_1)
+20\eta^{ab}\partial_a(ln{\Omega_1})\partial_b\Omega_1] \nonumber \\ 
+&2&\Omega_2^{-2}[\eta^{ab}\partial_a\partial_b(ln\Omega_1) 
-\eta^{ab}\partial_a\partial_b(ln\Omega_2)] 
\nonumber
\end{eqnarray}
Let us consider the case where
\begin{equation}
\Omega_1=\Omega_1(k_1y_1+k_2y_2)~~~,~~~~
\Omega_2=\Omega_2(k_3y_1+k_4y_2)
~~~~~~~~~y_1=x_4~,~y_2=x_5  \label{a14}
\end{equation}
Then the curvature scalar is obtained as
\begin{eqnarray}
R\,=
\Omega_1^{-2}g^{\mu\nu}R_{\mu\nu}
&-&\Omega_1^{-2}(k_1^2-k_2^2)
[10\frac{d^2(ln\Omega_1)}{du_1^2}+
20(\frac{d\,ln\Omega_1}{du_1})^2] \nonumber \\
&+&2\Omega_2^{-2}(k_1^2-k_2^2)
\frac{d^2(ln\Omega_2)}{du_1^2} 
-2\Omega_2^{-2}(k_3^2-k_4^2)
\frac{d^2(ln\Omega_2)}{du_2^2} 
\label{a15} \\
&&~~~~~~~u_1=k_1y_1+k_2y_2
~~~~~~~u_2=k_3y_1+k_4y_2
\nonumber
\end{eqnarray}
Because $k_1$, $k_2$ are projected out by $\partial_1$, $\partial_2$ 
they transform under (\ref{a1a}) 
like $\partial_1$, $\partial_2$; respectively
\begin{equation}
k_a\,\rightarrow\,-i\,k_a~~~~\mbox{as}~~~~
x_a \rightarrow i\, x_a~~~~~~
a=1,2  \label{a17a}
\end{equation}
So $k_1y_1+k_2y_2$ is automatically invariant under (\ref{a1a}) hence 
$\Omega_1$, $\Omega_2$ automatically obey (\ref{a1b}).
The application of the requirement, Eq.(\ref{a10}) to $\Omega_1$ and 
$\Omega_2$ results in
\begin{equation}
k_1=k_2~~~~,~~~~~k_3=k_4  \label{a17}
\end{equation}  
So the extra dimensional contribution to the curvature scalar i.e. $R_2$ 
in (\ref{a8}) vanishes. In other words the symmetry (\ref{a10}-\ref{a11}) 
requires 
the contribution to the cosmological constant from the extra 
dimensional curvature scalar be zero as well. 	   
The metric given in (\ref{a14}) is a smooth function of $x_4$, $x_5$. So 
the form of metric and the fact 
one of the extra dimensions is space-like and the other is space-like 
\cite{Li}
brings the over-all factors of 
$k_{1(3)}^2-k_{2(4)}^2$ which 
vanish in the limit of the symmetry (\ref{a11}) to make the 
curvature scalar zero. However if the metric tensor is not a smooth 
function of $x_A$ then $R_2$ does not have 
$k_{1(3)}^2-k_{2(4)}^2$ 
as over-all
factors however $R_2$ is still zero. 
To be more specific 
we consider a metric of the form of (\ref{a12}) and (\ref{a14}) with
\begin{equation}
\Omega_1^2=cos(|k_1y_1||+|k_2y_2|)
~~~~~~~~~
\Omega_2^2=0~~~~~~~~~y_1=x_4~,~y_2=x_5  
\label{a18}
\end{equation}
where a $Z_2\times\,Z_2$ orbifold symmetry induced by 
$k_1y_1\rightarrow\,-k_1y_1$
$k_2y_2\rightarrow\,-k_2y_2$
to get the 
absolute value signs in (\ref{a18}) and two branes located at 
the points $k_1y_1=0$, $k_2y_2=0$, $k_1y_1=\pi$, $k_2y_2=\pi$. By using 
Eq.(\ref{a15}) 
we obtain the curvature scalar as
\begin{eqnarray}
R=
\frac{1}{cos(|k_1y_1||+|k_2y_2|)}
&[&g^{\mu\nu}R_{\mu\nu}+10\,k_1^2\,tan\,u\;\tilde{\delta_1}
-10\,k_2^2\,tan\,u\;\tilde{\delta_2} 
+5(k_1^2-k_2^2)] \label{a19} \\
&&~~~~~~~u=|k_1y_1|+|k_2y_2|
\nonumber
\end{eqnarray}
where $\tilde{\delta_1}
= \delta(k_1y_1)-\delta(k_1y_1-\pi)$, $\tilde{\delta_2}=
\delta(k_2y_2)-\delta(k_2y_2-\pi)$ and 
$g^{\mu\nu}R_{\mu\nu}$ depends only on $x_\mu$.
Each delta function defines a 
5-dimensional subspace
and the intersections of these 5-dimensional subspaces define four 
3-branes which consist 
of two pairs of 3-branes related by the reversal of their signatures.
We see that $R_2$ due to Eq.(\ref{a19}) is zero in this case as well when 
$k_1=k_2$ (i.e. when the symmetry in Eq.(\ref{a11}) is exact) after $R$ 
is integrated over $y_1$ and $y_2$.

One must break this symmetry by a small amount in order to get a 
small cosmological constant in agreement with observations.
First we consider the metric (\ref{a18}). Assume that the symmetry imposed 
by (\ref{a11}) is broken for the metric (\ref{a18}) by a small amount 
$k_1-k_2=\Delta$. Then the 4-dimensional cosmological constant induced by 
$R_2$ is approximately equal to
\begin{equation}
\frac{2k_1\Delta}{16\pi\,G}\int\,5\,cos^2u 
\,dy_1dy_2\; =\;
\frac{5\pi\,k_1\Delta}{4G|k_1|^2}
\label{a20}
\end{equation}
We notice that the induced cosmological constant 
\begin{equation}
\Lambda^{(4)}= 
\frac{20k_1\pi^2\Delta}{|k_1|^2} \label{a21}
\end{equation}
will be even closer to zero if $k_1\approx k_2$ is large and the smallness 
of the cosmological constant is protected by the symmetry.
We have shown that the breaking of the symmetry  for the metric 
(\ref{a18}) leads to a 
a small cosmological constant provided the symmetry is broken 
by a small amount. In other words a small breaking of the symmetry 
does not lead to a big cosmological constant. Now we get a more general 
conclusion for the more general class of metrics (\ref{a9}).   
The Einstein equations corresponding to a 
conformally transformed metric $\Omega^2g_{\mu\nu}$ relate to the 
Einstein equations corresponding to the original metric in six 
dimensions as
\begin{equation}
\tilde{G}_{AB}=
G_{AB}
+4\delta_A^a\delta_B^b(
\partial_a\,ln\Omega
\partial_b\,ln\Omega
-\partial_a\partial_b\,ln\Omega)
+\tilde{g}_{AB}(6\eta^{ab}\partial_a\,ln\Omega\partial_a\,ln\Omega
+4\eta^{ab}\partial_a\partial_b\,ln\Omega) \label{a22}
\end{equation}
where
$\tilde{G}_{AB}=\tilde{R}_{AB}-\frac{1}{2}\tilde{g}_{AB}\tilde{R}$ is the 
Einstein tensor corresponding to $\tilde{g}_{AB}=\Omega^2g_{AB}$ and 
$G_{AB}=R_{AB}-\frac{1}{2}g_{AB}$ is the Einstein tensor corresponding to 
$g_{AB}$. 
The terms containing $\Omega$ on the right-hand side of (\ref{a22}) 
may be identified as the terms corresponding to the energy-momentum 
tensor induced by the conformal transformation. Meanwhile 
we observe that Dirac delta function can be written as 
\begin{equation}
\lim_{\alpha\rightarrow \infty} \alpha[1-tanh^2(\alpha z)]=\delta(z)
\label{a23}
\end{equation}
which follows from the fact that the derivative of step function gives the 
Dirac delta function. If we let 
$ln\Omega=\beta\,ln\,cosh\,\alpha(y_1-y_0)$ then the non-vanishing 
terms in 
Eq.(\ref{a22}) give
\begin{equation}
\partial_1\,ln\Omega\partial_1\,ln\Omega\,\rightarrow 
\,\beta^2\,\alpha^2
~~~~\mbox{and}~~~~
\partial_1\partial_1\,ln\Omega \rightarrow 
\beta\,\delta(y_1-y_{01})~~
~~~~~\mbox{as}~~~\alpha\rightarrow \infty
\label{a24}
\end{equation}
A small $\beta$ corresponds to the breaking of the symmetry by a small 
amount. If we take 
$\beta=\epsilon\frac{1}{\alpha}$ where $\epsilon\,<<\,1$ then a small 
perturbation in energy-momentum distribution leads to an even 
smaller bulk cosmological constant and results in an over-all rescaling 
of the metric by $\Omega$. 

The restriction put on the form of the matter Lagrangian by the 
symmetry Eq.(\ref{a1b}) can be determined 
by requiring the invariance of the corresponding action which requires 
the Lagrangian ${\cal L}$ transform as, ${\cal L}\rightarrow 
(-i)^D\,{\cal L}$. Then the transformation rule for the scalar field 
follows as
\begin{equation}
\frac{1}{2}g^{AB}\partial_A\phi\partial_B\phi\,\rightarrow
(-i)^D\frac{1}{2}g^{AB}\partial_A\phi\partial_B\phi~~~\mbox{implies}~~~~
\phi\,\rightarrow\,(-i)^{\frac{(D-2)}{2}}\phi \label{a20a}
\end{equation}
The mass term, $\frac{1}{2}m^2\phi^2$ is compatible with this symmetry 
since 
$m^2\rightarrow\,-m^2$ (which follows from $p^2=m^2$). However a $\phi^4$ 
term is not compatible with this symmetry (unless $D=4$). So this 
symmetry implies that $\phi^4$ terms may be induced only on 
4 dimensional branes. This together with the zero ( or almost zero) value 
of cosmological constant requires a two branes ( or even number of 
branes) scenario where $\phi^4$ 
terms are induced at both of the branes simultaneously and their 
contribution cancel ( or almost cancel) after integrated out over the 
extra dimensions. The transformation rule for fermions follows as 
\begin{equation}
i\bar{\psi}\gamma^{A}\partial_A\psi\,\rightarrow\,(-i)^D
i\bar{\psi}\gamma^{A}\partial_A\psi 
~~~\mbox{implies}~~~~\psi\,\rightarrow\,(-i)^{\frac{(D-1)}{2}}\psi
\label{a20b}
\end{equation}
The mass term 
$m\bar{\psi}\psi$, the fermion-scalar interaction term 
and the fermion - gauge boson interaction term 
$i\bar{\psi}\gamma^{A}B_A\psi$, all are compatible with the symmetry.
while 
\begin{equation}
\bar{\psi}\psi\phi\,\rightarrow\, (-i)^\frac{3D-4}{2} \label{a20ba}
\end{equation}
is compatible with the symmetry only for $D=4$. So this term may only be 
induced on a 4-dimensional brane.
Because $\partial_A$ and $B_A$ are combined in the covariant derivative  
$\partial_A-iB_A$, $B_A$ must transform in the same way as $\partial_A$.
This implies that the gauge field kinetic term $-\frac{1}{4}F_{AB}F^{AB}$ 
\begin{equation}
F_{AC}F^{AC}=(\partial_A\,B_C-\partial_C\,B_A)
(\partial^A\,B^C-\partial^C\,B^A)\,\rightarrow\,  
(-i)^{D}F_{AC}F^{AC} \label{a20c}
\end{equation}
is also compatible with the symmetry. 

A comment is in order at this point. 
We have found in the above paragraph that the mass terms are allowed in 
all dimensions unlike 
the usual scale invariant field theories and the result of 
Nobbenhuis \cite{nobbenhuis} although he uses the same symmetry 
as the one given here. The difference between the conclusions 
come from the difference in the identification of how mass terms behave 
under the scaling transformation.  In the usual scale invariance and in 
the Nobbenhuis's study \cite{nobbenhuis} masses are taken to be 
invariant under the symmetry transformation while in this study 
the masses transform like momenta ($p^2=m^2$). In fact the approach in 
the present study is in agreement with 
the identification of mass terms as 
the kinetic terms of the higher dimensions 
and under this condition this is the only 
consistent transformation provided that one scales all the 
dimensions simultaneously. Otherwise it means that either one does not 
scale all the dimensions or does not consider the mass terms as 
the kinetic terms of higher dimensions. In fact the difference in the 
approach 
to the scaling property of mass term is just a matter of convenience. It 
depends on one's aim of using the scale invariance. If one just tries to 
get phenomenological results confined to relatively low energies where  
the extra dimensions related to the masses are not observable one should 
take the masses be invariant under scale transformation. However if one  
tries to get general results applicable to all dimensions ( as is the case 
in this study) one should transform the mass terms like momenta because 
it is just the kinetic term written in another form in this case.
Another difference between the result of Nobbenhuis and mine is that he 
reaches the conclusion that the cosmological constant must 
vanish in the usual four dimensions, $D=4$ while I obtain the same result 
for $D=2(2n+1)$ i.e $D=2,6,10$. This difference is due to different 
methods employed in the implementation of the symmetry. Nobbenhuis 
uses the covariance of the equations of motion ( i.e Einstein field 
equations) under the symmetry employed in this paper while I use 
the invariance of the action functional under the same symmetry as the 
basis of my arguments. The requirement, of the covariance of the Einstein 
equations, used in Nobbenhuis's study leads to the result that the 
Einstein tensor and cosmological 
constant can not coexist, either of them must vanish. This method does not 
tell anything about the allowed number of dimensions. The conclusion of 
Nobbenhuis depends on the assumption that the Einstein tensor is 
already allowed in four dimension, so the cosmological constant must 
vanish. On the other hand the requirement, 
of the invariance of the action, used in the present study does not only 
give the result that both 
of the cosmological constant and the Einstein tensor can not coexist but 
it also leads to the formula for the number of dimensions, 
$D=2(2n+1)=2,6,10,..$, which forbid a non-vanishing cosmological constant. 
In 
other words 
in our analysis the Ricci scalar can not be four dimensional in origin, it 
must be induced, as an effective four dimensional Ricci scalar, from 
higher dimensions ( e.g. from a 6 dimensional Ricci scalar) 
so that it has some hidden invariance under the symmetry ( through the 
extra dimensional parameters which are integrated out).

Given the fact that the symmetry employed here is a sub-case of the 
complexified version of the scale invariance ( hence of the conformal 
invariance) and the fact that there 
are some serious problems with the quantization of the classical field 
theories with conformal invariance \cite{conformal} ( especially in 
dimensions other than two ) one may wonder if a similar problem exists for 
this symmetry, 
that is, if the symmetry introduced in this paper is preserved after 
quantization. It is a well-known fact that the lack of persistence of the 
scale invariance in quantum field theory results from the non-invariance 
of the correlation functions ( i.e n-point functions) under scale 
invariance. This non-invariance results in the nonconservation of the 
Noether current due to the breakdown of the scale symmetry 
after the quantization. 
This, in turn, induces conformal anomalies which are rather difficult to 
manage (especially in dimensions other than two). This situation is 
improved in the case of the symmetry introduced in this paper. As it is 
evident from Eqs.(\ref{a20a}, \ref{a20b}, \ref{a20c}) the 2-point 
functions ( which serve as the building blocks for n-point functions) 
scale as $(-i)^{(D-2)}$ for scalars and gauge fields and scale as 
$(-i)^{(D-1)}$ for fermions. Therefore the 2-point functions are invariant 
under this symmetry in the dimensions $D-2=4n$ ($D=4n+2$) for scalars and 
gauge fields and in $D-1=4n$ ($D=4n+1$) for fermions (where 
$n=0,1,2,....$). Hence one may speculate models where the 
renormalizability of the model for scalars-gauge fields and fermions is 
manifest at different dimensions higher than four ( e.g, say at $D=11$ or 
$D=7$ for fermions and at $D=10$ or $D=6$ for scalars and gauge fields) and 
at lower dimensions the theory behaves as an effective theory with a ( in 
some sense) hidden symmetry. For the time being, these remarks are just 
speculations . A detailed study of this topic is necessary to 
arrive reliable conclusions about this point. In any case, 
with respect to quantization,
this symmetry seems to be more promising than the usual scale symmetry.

I hope that this study will give additional insight towards the solution 
of the cosmological constant problem. However there is more work to be 
done in this direction. The metric employed here is static so the 
cosmological constant is constant in time. The need for inflation in the 
history of the universe needs a much larger value for the cosmological 
constant in the early universe. So the next step should be making the 
metric time dependent to get a time dependent cosmological constant. The 
self-tuning solutions with large extra dimensions discussed in the 
introduction \cite{burgess} need smaller sizes for the extra dimensions in 
the inflationary universe era to get larger cosmological constants in that 
era. On the other hand the extra dimensions here may be small or large. 
Moreover the extra dimensions in the inflationary era may be larger or 
smaller than their present values. For example a term of $k_0t$ in the 
argument of the conformal term $cosu$ may break the symmetry thus 
induce a cosmological constant. The induced cosmological term would depend 
on the phase factor $k_0t$ hence it may be different at different times 
independent of the size of the extra dimensions. Another point which needs 
further study is the source of this symmetry and its breaking 
mechanisms both in a physical and 
a mathematical content. I hope the investigation of all these and 
other interesting unanticipated aspects of this symmetry give fruitful 
results.

\begin{acknowledgments}
I would like to thank Prof. Durmu{\c{s}} Ali Demir for reading 
the manuscript and fruitful discussions on the topic and for his valuable 
comments. 
\end{acknowledgments}



\bibliographystyle{plain}

\begin{thebibliography}{99}
\bibitem{nobbenhuis} After the submission of this work, hep-th/0410063 
to the ArXive I have learnt from the paper given below that this 
symmetry principle was already noticed by G. 't Hooft ( though 
unpublished)\\ 
S. Nobbenhuis, ArXive gr-qc/0411064
\end{thebibliography}

\end{document}